\documentclass[twoside,slac_one]{revtex4}
\usepackage{graphicx}
\usepackage{fancyhdr}
\usepackage{amsmath} 
\usepackage{bm}
\usepackage{amsxtra}
\usepackage{amssymb}
\usepackage{amsthm}
\usepackage{latexsym}
\usepackage{lscape}

\usepackage{xspace}
\usepackage{subfig}

\newcommand{\pt}{\ensuremath{p_{\mathrm{T}}}\xspace}

\newcommand{\kt}{\ensuremath{k_t}\xspace}
\newcommand{\met}{\ensuremath{\not\!\!{E_{T}}}\xspace}
\newcommand{\hT}{\ensuremath{H_{\mathrm{T}}}\xspace}

\newcommand{\fbinv} {\mbox{\ensuremath{\,\text{fb}^\text{$-$1}}}\xspace}

\newcommand{\gev}{\ensuremath{\,\text{Ge\hspace{-.08em}V}}\xspace}
\newcommand{\tev}{\ensuremath{\,\text{Te\hspace{-.08em}V}}\xspace}

\begin{document}

\title{Search for new physics with same-sign isolated dilepton events with jets and missing
  transverse energy at CMS}

%

\author{M. Weinberg}
\affiliation{Department of Physics, University of Wisconsin, Madison, WI, USA}

\begin{abstract}
The results of searches for Supersymmetry in events with two same-sign isolated leptons, hadronic
jets, and missing transverse energy in the final state are presented. The searches use $pp$
collisions at 7~\tev collected in 2011 by the CMS experiment.
\end{abstract}

\maketitle

\thispagestyle{fancy}


\section{Introduction}
Events containing isolated same-sign dileptons are very rare in the Standard Model (SM), but can
occur quite naturally in many different new physics models, including supersymmetry
(SUSY)~\cite{susyLikeSign} and universal extra dimensions~\cite{bosonicSusy}. Isolated same-sign
lepton pairs are thus a very clean experimental signature for new physics searches.

In addition, the analysis described in this talk requires significant missing transverse energy
($\met$) and hadronic activity in the form of high-$\pt$ jets. The choice of signal regions is
influenced by two experimental observations. First, astrophysical evidence for dark
matter~\cite{particleDarkMatter} suggests the need for a massive, weakly-interacting stable particle
which gives rise to final states with $\met$. Second, new physics signals with observably large
cross sections are likely to be the result of strong interactions, and we therefore expect them to
be accompanied by significant hadronic activity. Aside from these requirements, our searches are as
independent of the particular details of new physics models as possible.

The data used in this analysis were collected in $pp$ collisions at a center-of-mass energy of
7~\tev by the Compact Muon Solenoid (CMS) detector at the Large Hadron Collider (LHC) in 2011. They
comprise a total integrated luminosity of 0.98~\fbinv.

\subsection{Same-Sign Dileptons in SUSY}

\begin{figure}[ht]
\centering
\includegraphics[width = 80mm]{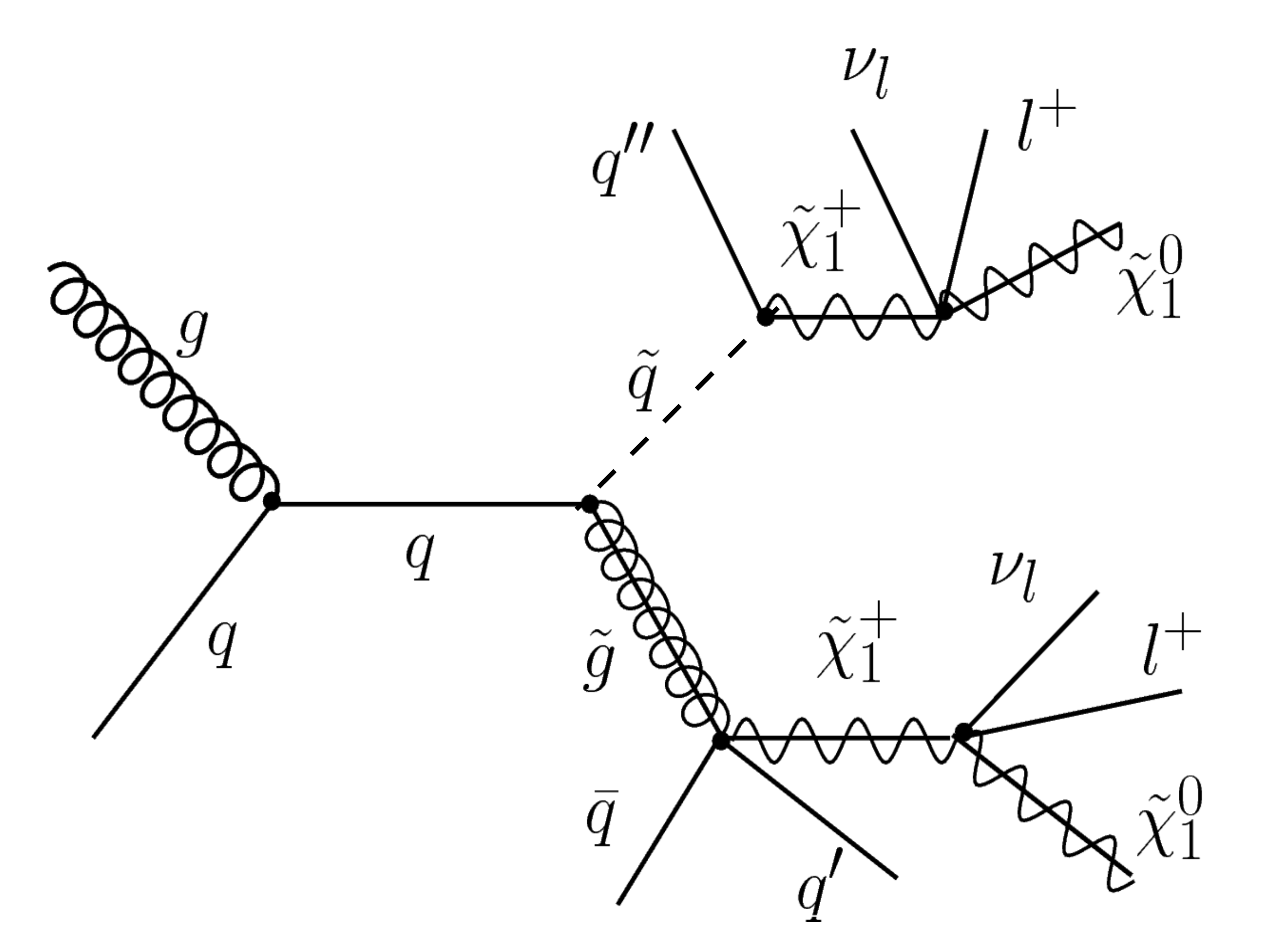}
\caption{An example of a process involving the production and decays of SUSY particles, which gives
  rise to two same-sign prompt leptons, jets, and missing transverse energy.} \label{figure1}
\end{figure}

An example SUSY cascade decay leading to jets,~\met, and same-sign dileptons is shown in
Figure~\ref{figure1}. The produced gluinos or squarks decay to charged gauginos, which subsequently
decay to the lightest supersymmetric particle (LSP) neutralino. The mass difference between the
gluinos/squarks and the charged gaugino, typically arbitrary, defines the amount of hadronic
activity expected in the event. The mass difference between the gaugino and neutralino influences
the lepton~\pt spectrum. Further, there is a range of scenarios where a large production asymmetry
exists between the $\tau$ lepton and the $e/\mu$ leptons. The range of mass differences motivates a
variety of selection criteria in order to cover the widest possible phase space, while the possible
tau production asymmetry motivates us to look specifically at events containing taus.

\section{Reconstruction of Leptons, Missing Energy, and Jets}

Electrons, muons, and hadronically decaying taus are all included in the analysis. Lepton candidates
are required to have $|\eta| < 2.4$ and to be consistent with originating from the same interaction
vertex.

Electron candidates~\cite{elecCand} consist of an energy cluster in the electromagnetic calorimeter
(ECAL) matched to hits in the tracker. The identification of electrons is based on the shape of
their electromagnetic shower in the ECAL as well as track-cluster matching. The criteria are
designed to maximally reject electron candidates from QCD multijet production, and are approximately
80\% efficient for electrons from the decay of $W/Z$ bosons.

Muon candidates~\cite{muonCand} must be reconstructed via two separate algorithms: Tracker muons are
seeded from hits in the tracker and are matched to signals in the calorimeters and muon systems.
Global muons are constructed from a simultaneous fit to hits in both the tracker and muon chambers.
The identification efficiency measured in data is approximately 96\% for muons of all momenta.

Jets and~\met are reconstructed via the particle flow (PF) technique~\cite{particleFlow}, which
takes signals in each detector component and reconstructs physics objects from them before running
jet clustering algorithms. The hadronic jets in this analysis use the anti-\kt clustering algorithm
with a cone of $\Delta R = 0.5$ in $\eta-\phi$ space.

Hadronic tau candidates~\cite{tauCand} are reconstructed starting from jets. A variable size cone of
$\Delta R < 5~\gev/\pt$ is defined around the leading track, and the $\tau$ decay products are
required to be confined within this cone.

%

\section{Baseline Selections}

The analysis starts from an initial selection of two same-sign leptons with $\pt > 5, 10, 15~\gev$
for muons, electrons, and taus, respectively, and $|\eta| < 2.4$. Two jets with $\pt > 40~\gev$ and
$|\eta| < 2.5$ are also required, and we further require $\hT > 80~\gev$ and $\met > 30~\gev$.

This initial selection is further divided into three baseline selections: The \textit{inclusive}
selection for the $ee$, $e\mu$ and $\mu\mu$ final states requires $\hT > 200~\gev$. The
\textit{high-\pt} selection for the $ee$, $e\mu$ and $\mu\mu$ final states requires $\pt(l_{1},
l_{2}) > 20, 10~\gev$. The tau-specific selection requires at least one of the final state particles
is a tau ($e\tau$, $\mu\tau$, or $\tau\tau$) with $\hT > 350~\gev$ and $\met > 80~\gev$.




\subsection{Search Regions}

We constrain the baseline selection regions with the following search regions:

\begin{itemize}
  \item \textit{High-\hT, high-\met}: $\hT > 400~\gev$ and $\met > 120~\gev$, providing a high
    expected sensitivity to points in the Constrained Minimal Supersymmetric Standard Model
    (CMSSM)~\cite{constrainedMinSusy} with low values of $m_{0}$.
  \item \textit{Medium-\hT, high-\met}: $\hT > 200~\gev$ and $\met > 120~\gev$, targeting models
    with moderate mass-splittings between squarks/gluinos and gauginos.
  \item \textit{High-\hT, low-\met}: $\hT > 400~\gev$ and $\met > 50~\gev$, providing a high
    expected sensitivity to CMSSM parameter points with high values of $m_{0}$.
  \item \textit{Low-\hT, high-\met}: $\hT > 80~\gev$ and $\met > 100~\gev$, providing a high
    expected sensitivity to models predicting low hadronic activity with high~\met.
\end{itemize}

\begin{figure*}[hbtp]
\centering
\subfloat[][]{
  \includegraphics[width = 50mm]{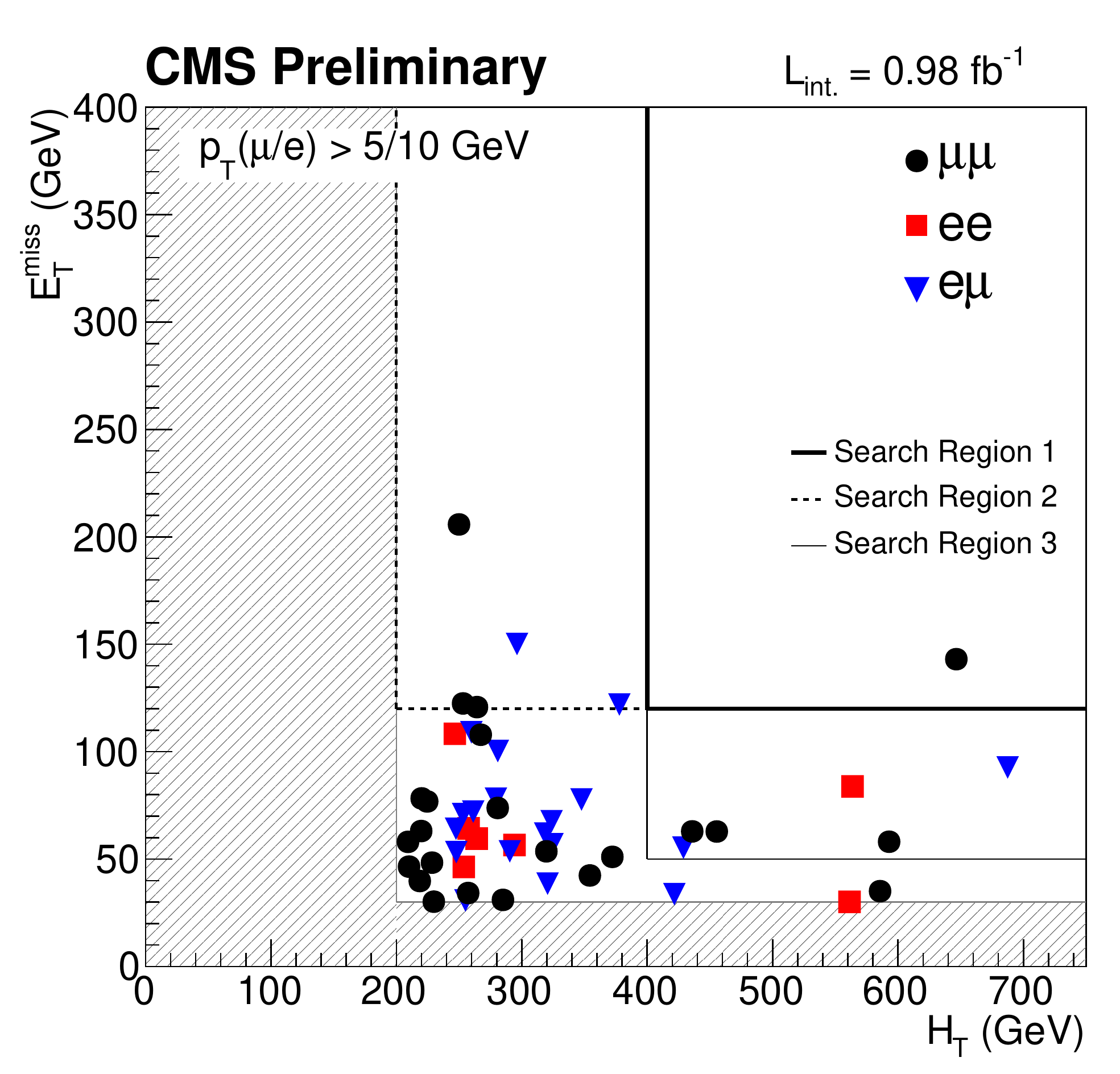}
  \label{figure2a}
}
\subfloat[][]{
  \includegraphics[width = 50mm]{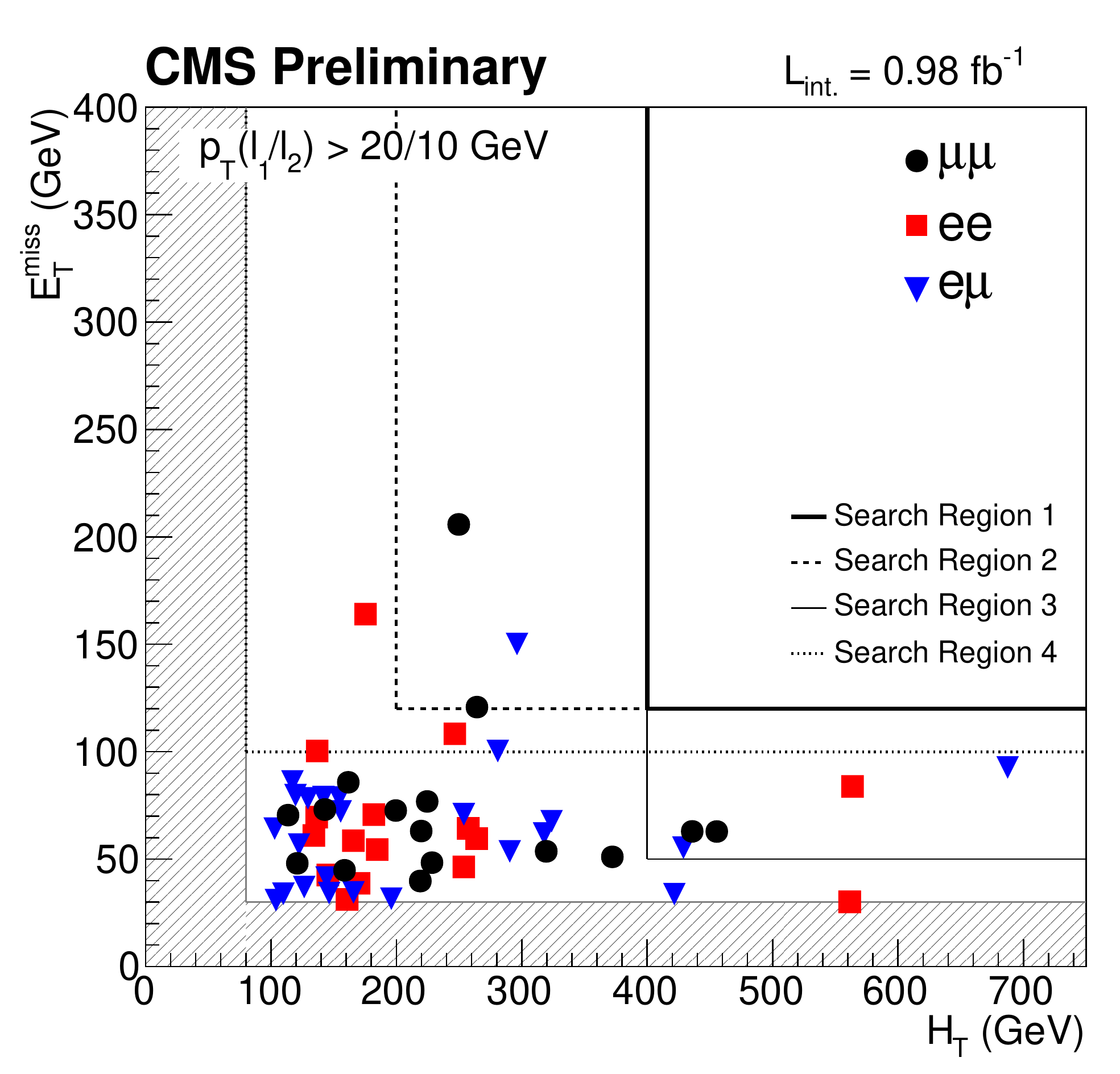}
  \label{figure2b}
}
\subfloat[][]{
  \includegraphics[width = 50mm]{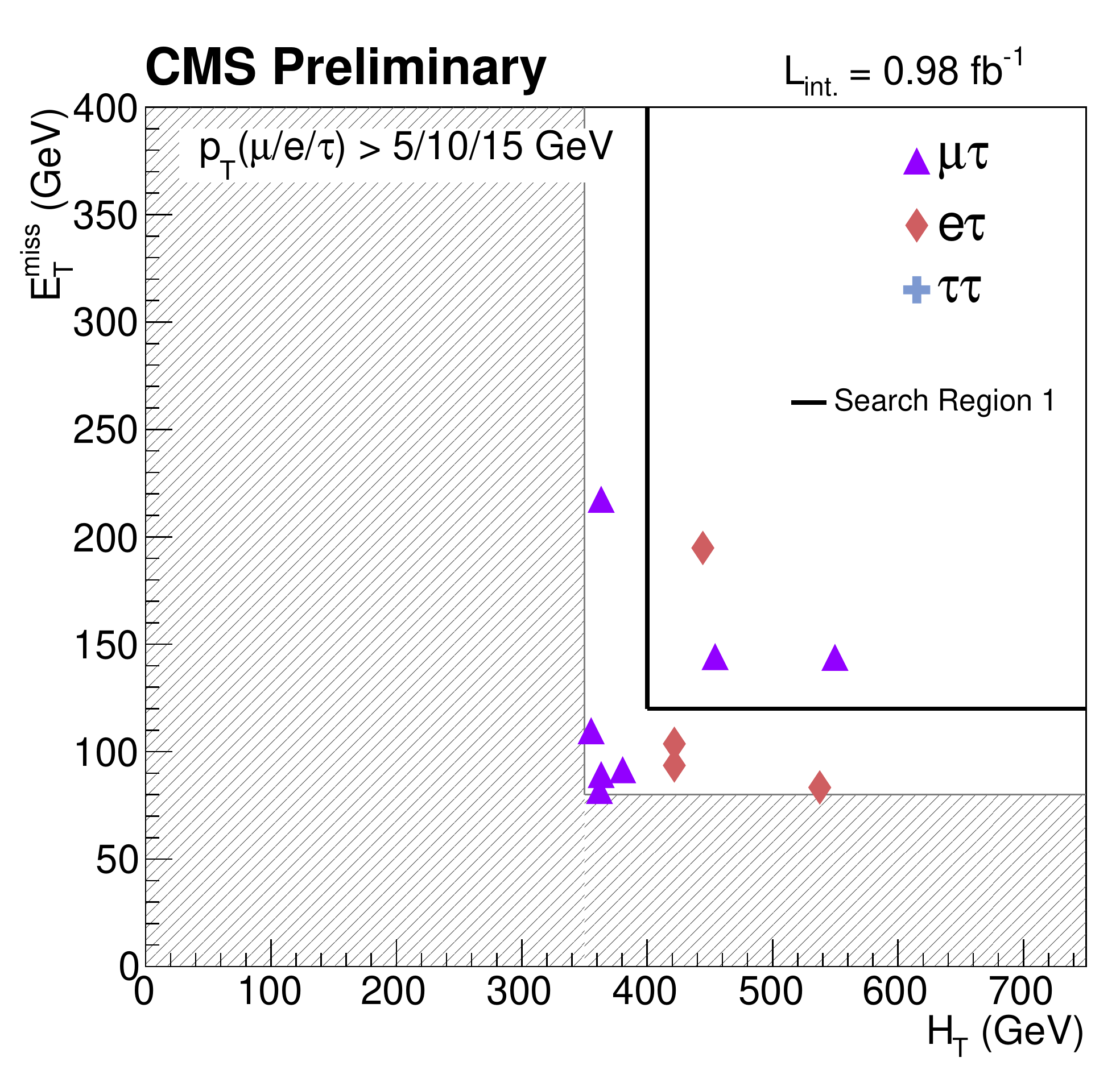}
  \label{figure2c}
}
\caption{\hT versus~\met scatter plots for the three baseline regions in data: inclusive
  dileptons~\subref{figure2a}, high-\pt dileptons~\subref{figure2b}, and
  $\tau$-dileptons~\subref{figure2c}.}
\label{fig:figure2}
\end{figure*}

Figure~\ref{fig:figure2} shows the events observed in data on the~\hT-\met plane for each baseline
selection categories, with the dashed and dotted lines indicating the various search regions.

\subsection{Background Estimation}

Backgrounds are divided into categories based on how many prompt leptons they contain. Rare
processes like $q\bar{q} \rightarrow WZ$ and $ZZ$, $qq \rightarrow qqW^{\pm}W^{\pm}$, $2 \times
(q\bar{q} \rightarrow W)$, and $t\bar{t}W$ can produce actual prompt same-sign leptons. These are
evaluated from Monte Carlo (MC) simulation and found to contribute between 10\% and 40\% of the
total background. A 50\% systematic uncertainty is applied to the value.

Processes such as $Z/\gamma^{*} \rightarrow l^{+}l^{-}$ and $t\bar{t}$ produce opposite-sign prompt
leptons. In cases where the charge of one of the leptons is misreconstructed, these events
constitute an additional background to the search. They are evaluated in data and found to
contribute less than 10\% to the total background.

\begin{figure*}[hbtp]
\centering
\subfloat[][]{
  \includegraphics[width = 80mm]{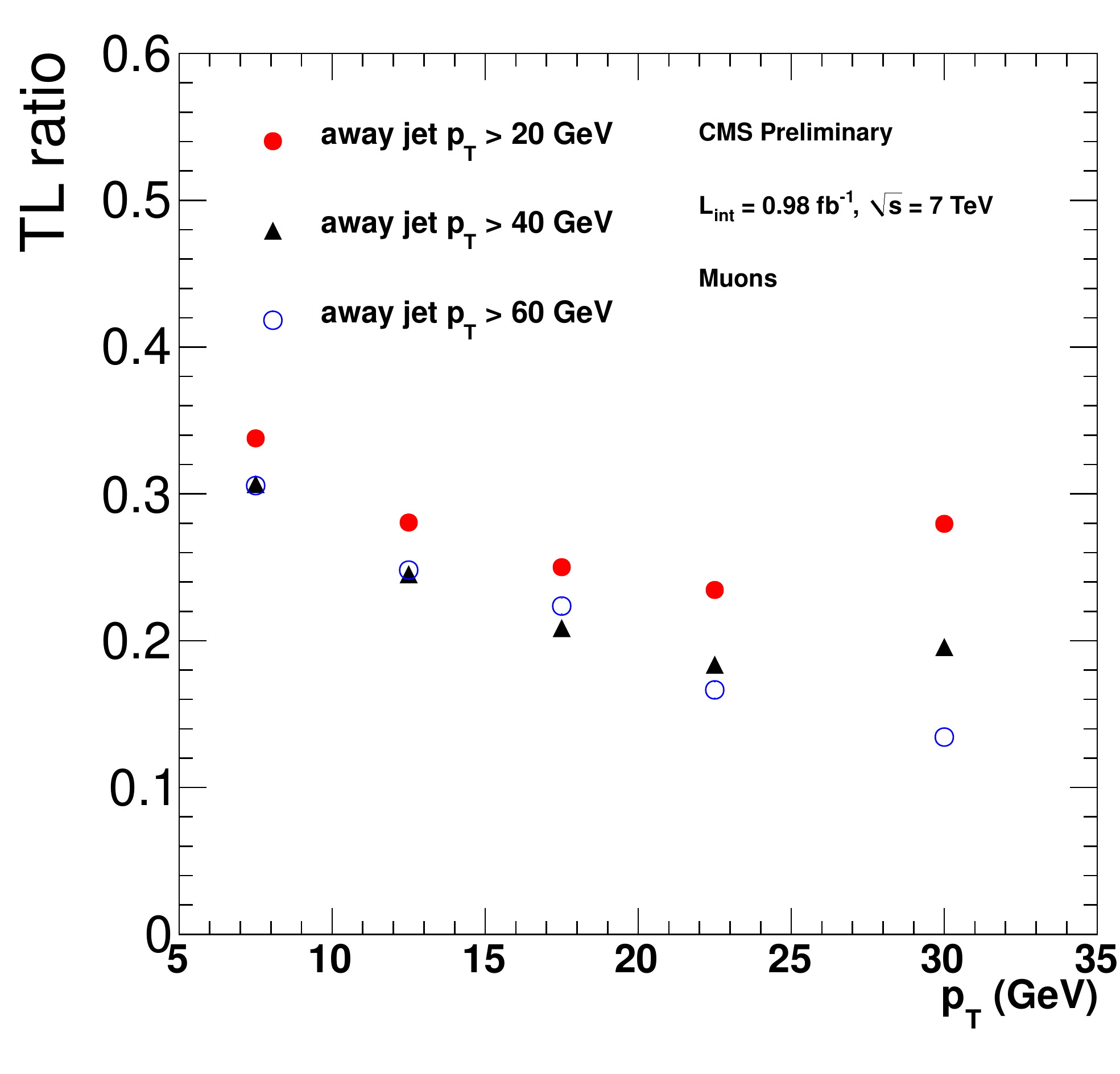}
  \label{figure10a}
}
\subfloat[][]{
  \includegraphics[width = 80mm]{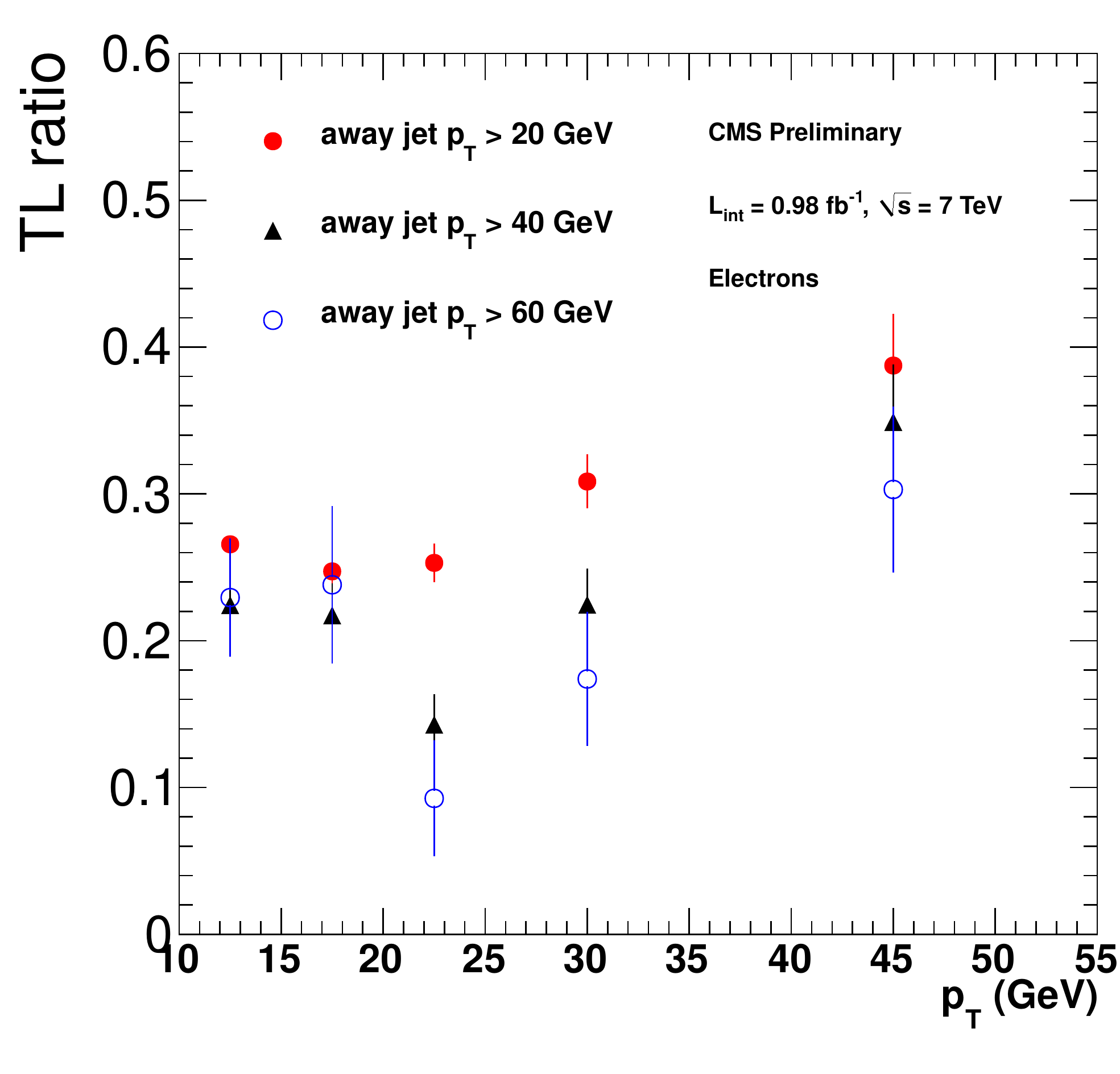}
  \label{figure10b}
}
\caption{Muon~\subref{figure10a} and electron~\subref{figure10b} TL ratio projected on~\pt using
  method (A1). The TL ratio distribution is displayed separately for events with a jet separated
  from the lepton candidate by $\Delta R > 1$ and the jet required to have~\pt above 20, 40, and
  60~\gev. The central value is measured for the case with a jet $\pt > 40~\gev$, while the range of
  values measured with other jet requirements represents an estimate of the systematic uncertainty.}
\label{fig:figure10}
\end{figure*}

Events with ``fake'' leptons from jets also contribute. Here, a fake lepton can be a lepton produced
in a heavy flavor quark decay as well as a light jet that is misidentified as a lepton. This is also
evaluated in data, and is found to be the dominant effect. In order to evaluate this fake rate, the
analysis loosens the lepton selection criteria (such as lepton isolation requirements) to define a
set of ``fakeable objects''. The ratio of lepton candidates passing the full selection
(\textit{tight} leptons) to lepton candidates passing this looser selection but failing the full
selection (\textit{loose} leptons) is then determined for QCD multijet events with a jet above a
given threshold. This tight-to-loose (TL) ratio is shown for both electrons and muons in
Figure~\ref{fig:figure10}.

\begin{figure*}[hbtp]
\centering
\subfloat[][]{
  \includegraphics[width = 50mm]{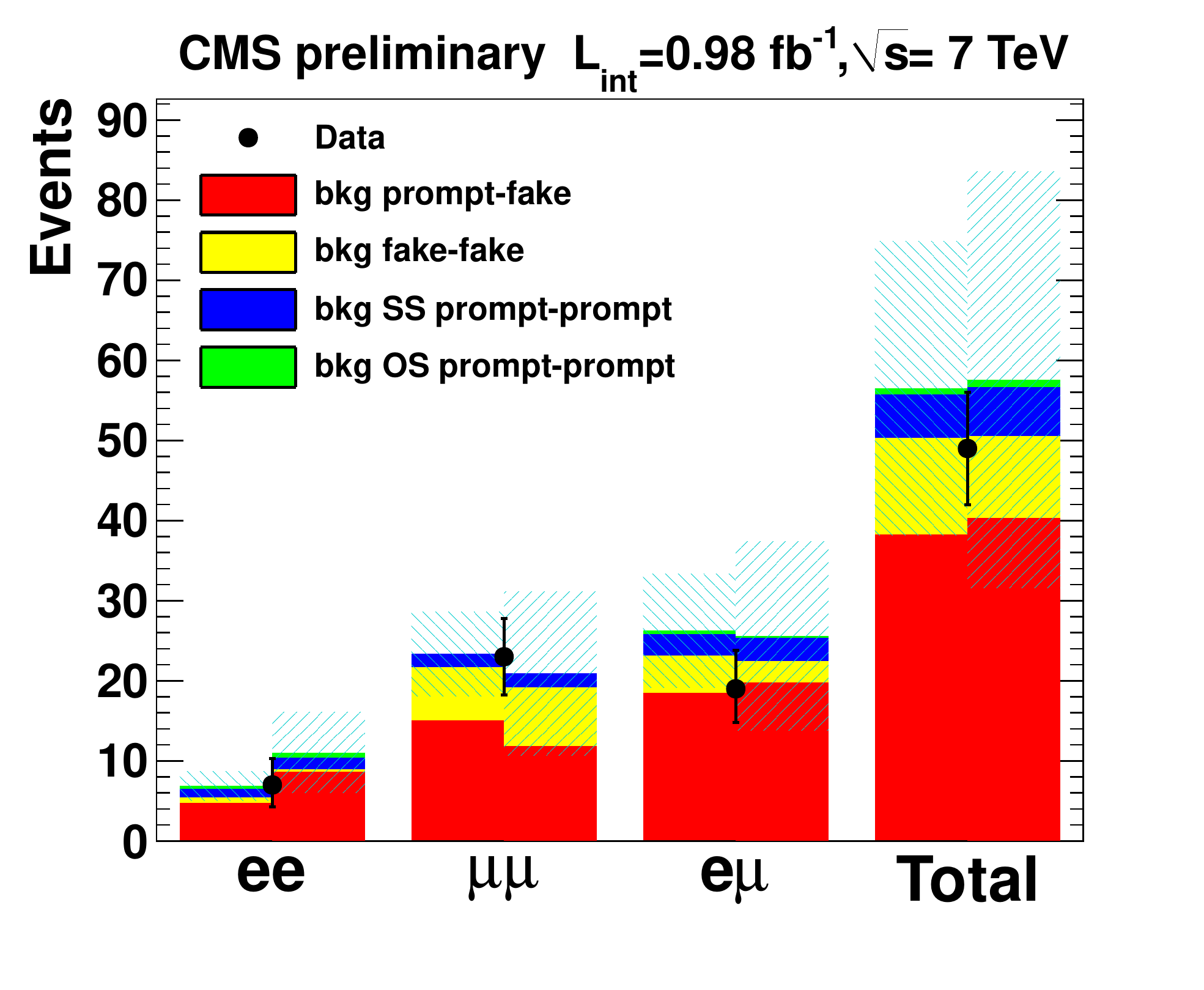}
  \label{figure3a}
}
\subfloat[][]{
  \includegraphics[width = 50mm]{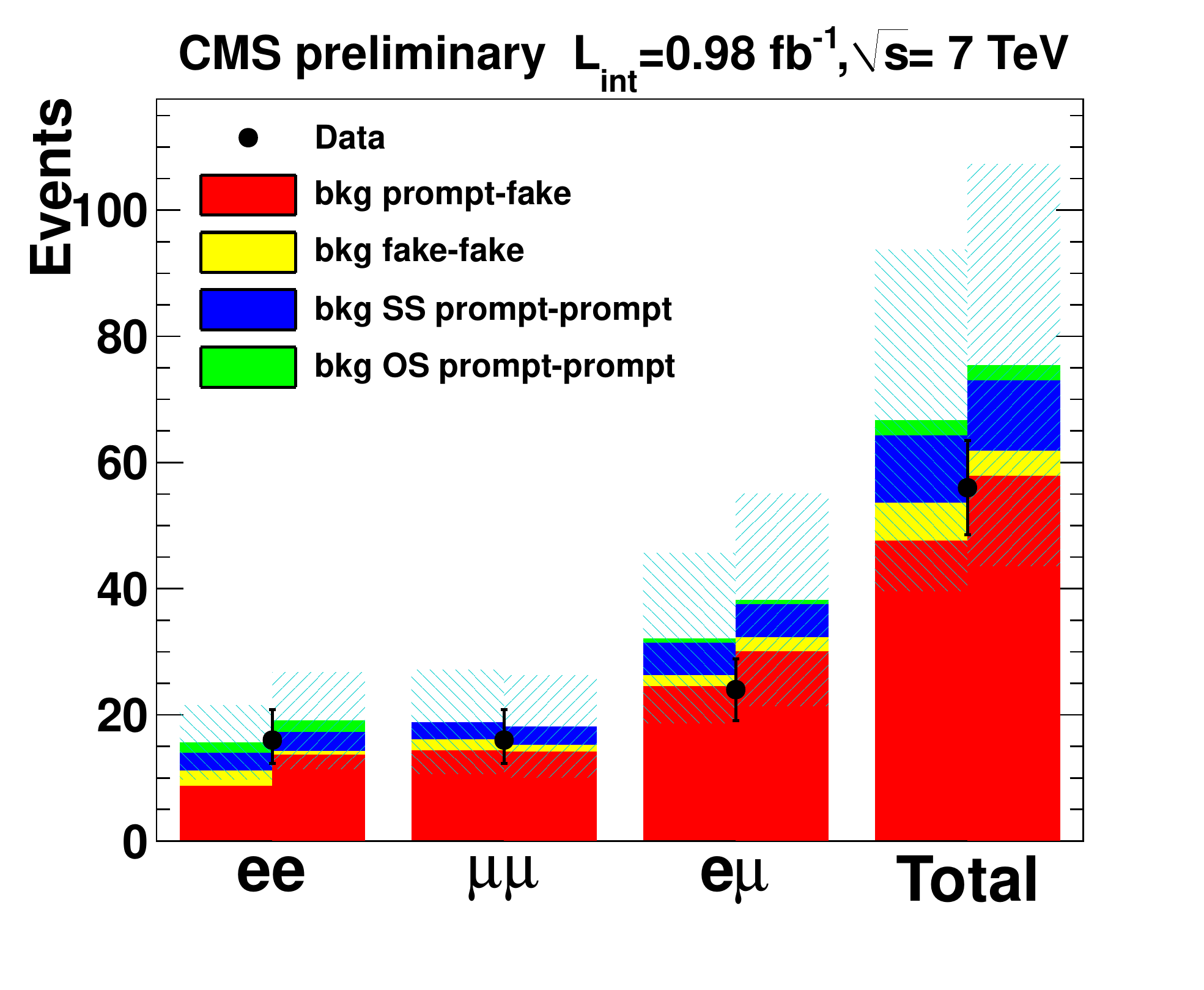}
  \label{figure3b}
}
\subfloat[][]{
  \includegraphics[width = 50mm]{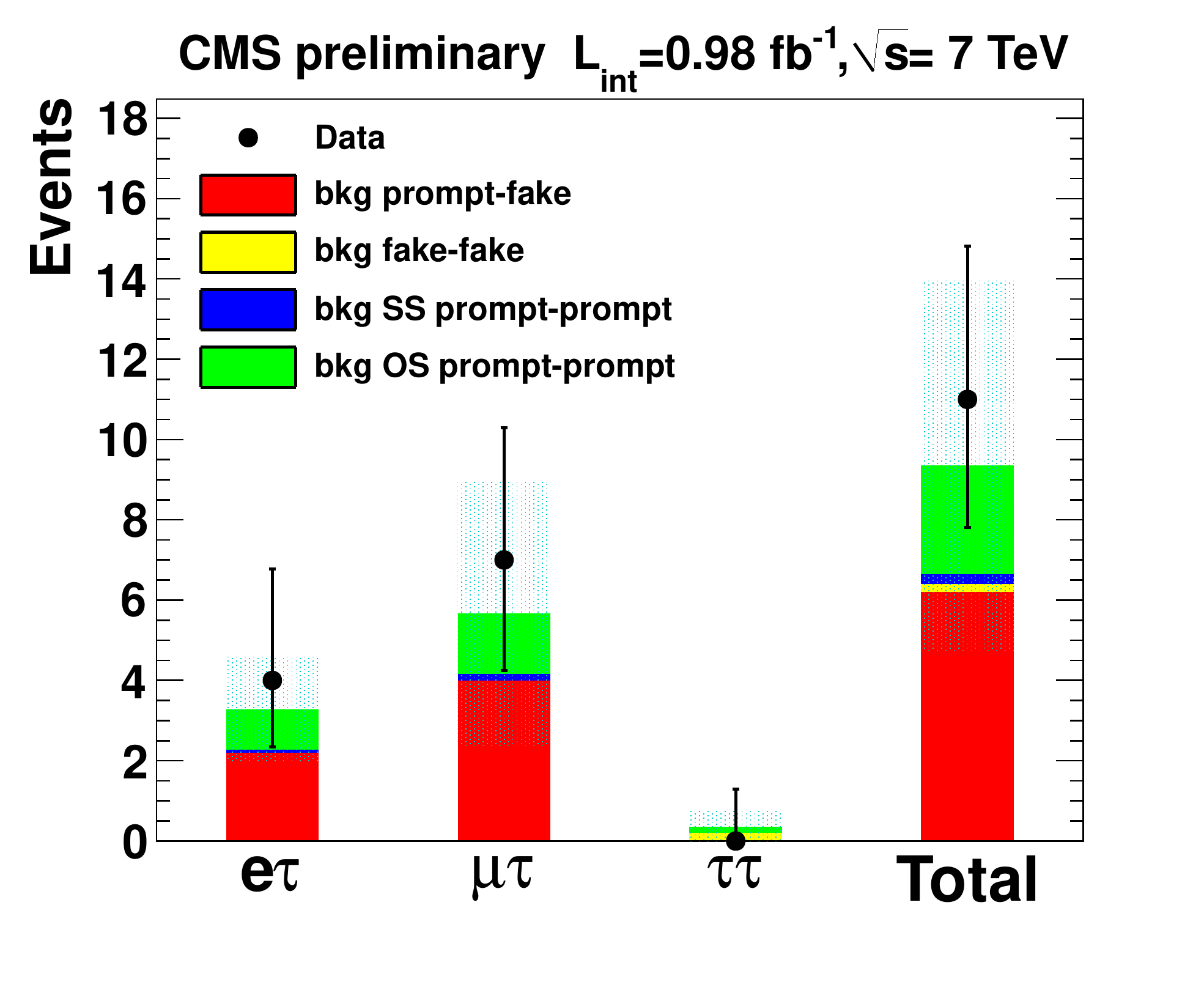}
  \label{figure3c}
}
\caption{Summary of background predictions and observed yields in the baseline region for the
  inclusive~\subref{figure3a}, high-\pt~\subref{figure3b}, and $\tau$
  dilepton~\subref{figure3c} selections. For the inclusive selections, the results of method (B) are
  compared with those from method (A1) in the left and right bar for each channel, respectively. For
  the high-\pt selections, the results of method (A2) are compared with those from method (A1) in
  the left and right bar for each channel, respectively. Predictions for events with one and two
  fakes (prompt-fake and fake-fake), contributions from simulated backgrounds (SS prompt-prompt),
  and those from events with a lepton charge misreconstruction (OS prompt-prompt) are reported
  separately.}
\label{fig:figure3}
\end{figure*}

A summary of the background predictions as well as the number of observed events is shown for each
channel in the three baseline selection regions in Figure~\ref{fig:figure3}. For the inclusive and
high-\pt dilepton selections, two sets of complementary methods are used to measure the backgrounds,
as indicated by the two bars associated with each channel. These methods compare well with each
other, providing mutually consistent background predictions.

\section{Search Results}

Table~\ref{table1} shows the events observed for each selection region, along with the total
background predictions. We see no evidence of an event yield in excess of the background predictions
and set 95\% CL upper limits on the number of observed events using a hybrid frequentist-bayesian CL
method~\cite{confidenceLimit} with nuisance parameters and the signal strengh maximizing the ratio
of the signal-with-background to background-only likelihoods. These limits are indicated in the
bottom row of the table.

\begin{table*}[t]
\begin{center}
\caption{Observed number of events in data compared to the predicted background yields for the
  inclusive, high-\pt, and $\tau$ dilepton search regions. The uncertainties include the statistical
  and systematic components added in quadrature. The last row (95\% CL UL yield) represents observed
  upper limits on event yields from new physics.}
\begin{tabular}{|c|c|c|c|c|c|c|c|c|}
\hline
\textbf{Dilepton category} & \multicolumn{3}{c|}{\textbf{Inclusive ($\hT > 200~\gev$)}} & \multicolumn{4}{c|}{\textbf{High-\pt~$\left[\pt(l_{1}, l_{2}) > 20, 10~\gev\right]$}} & \textbf{Taus} \\
\hline
\textbf{\hT(\gev)/\met(\gev)} & \textbf{400/120} & \textbf{400/50} & \textbf{200/120} & \textbf{400/120} & \textbf{400/50} & \textbf{200/120} & \textbf{80/100} & \textbf{400/120} \\
\hline
\textbf{Predicted} & $2.3 \pm 1.2$ & $5.3 \pm 2.4$ & $6.6 \pm 2.9$ & $1.4 \pm 0.7$ & $4.0 \pm 1.7$ & $4.5 \pm 1.9$ & $10 \pm 4$ & $2.9 \pm 1.7$ \\
\hline
\textbf{Observed} & 1 & 7 & 6 & 0 & 5 & 3 & 7 & 3 \\
\hline
\textbf{95\% CL UL yield} & 3.7 & 8.9 & 7.3 & 3.0 & 7.5 & 5.2 & 6.0 & 5.8 \\
\hline
\end{tabular}
\label{table1}
\end{center}
\end{table*}

\begin{figure*}[hbtp]
\centering
\subfloat[][]{
  \includegraphics[width = 80mm]{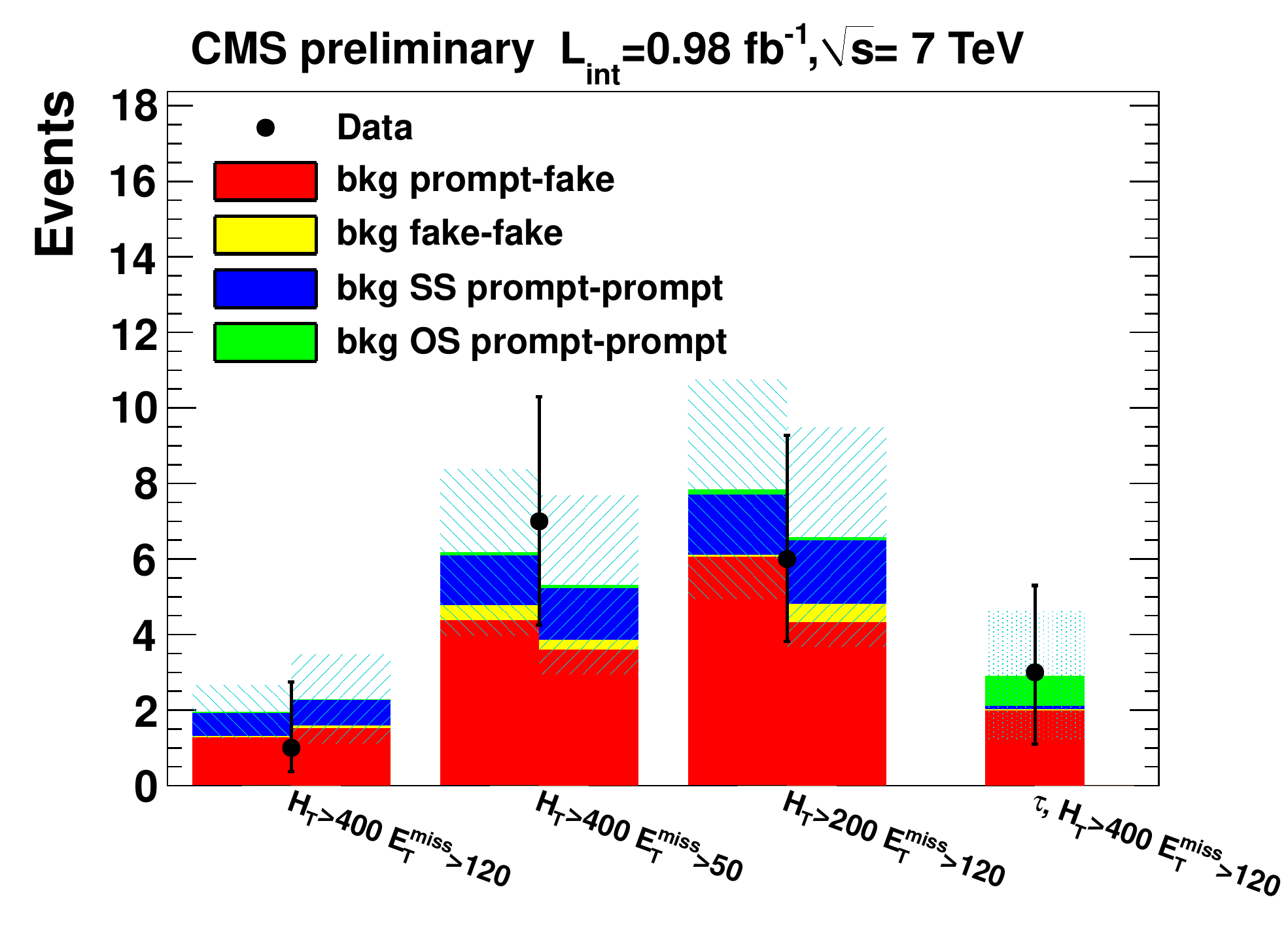}
  \label{figure4a}
}
\subfloat[][]{
  \includegraphics[width = 80mm]{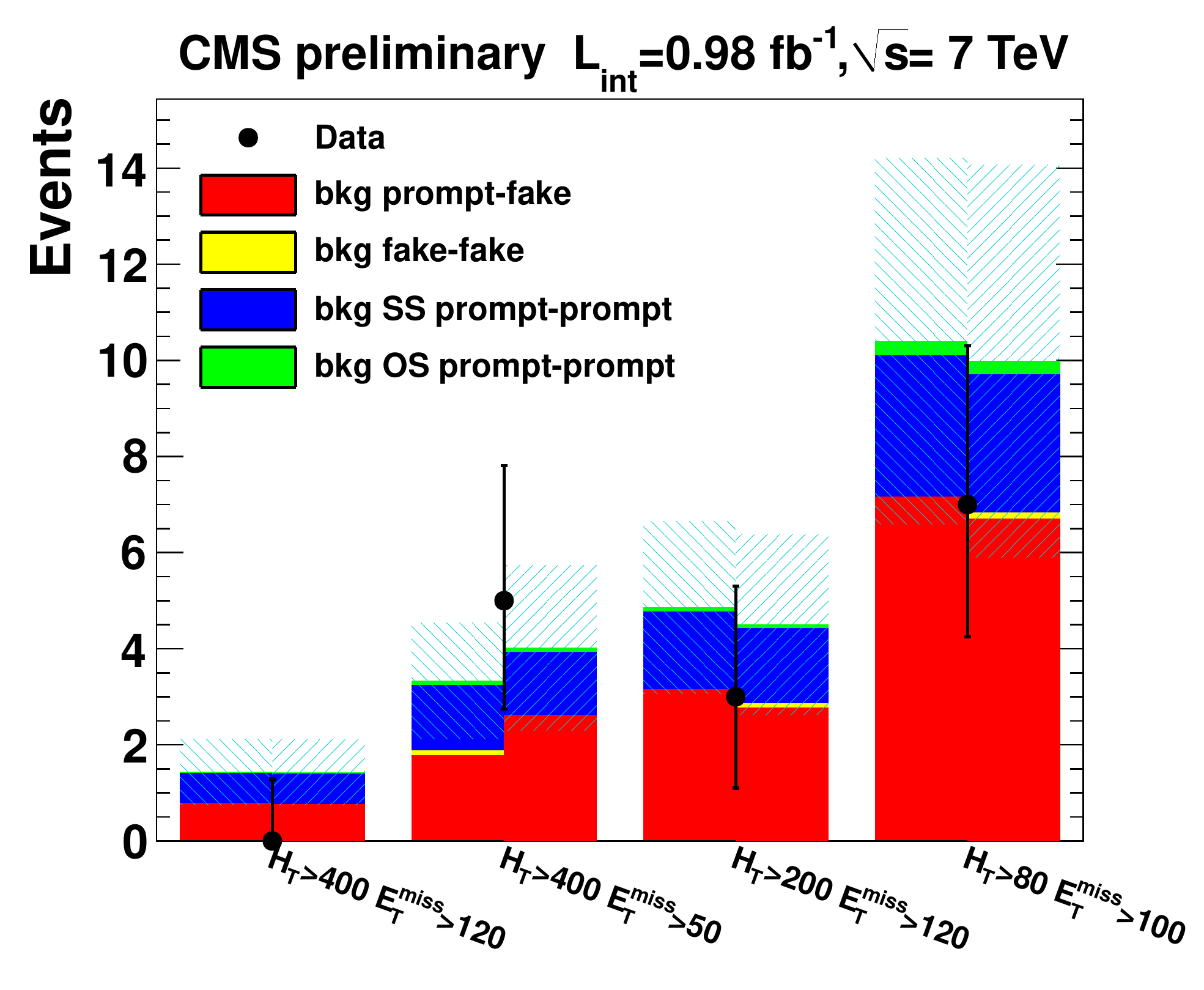}
  \label{figure4b}
}
\caption{Summary of background predictions and observed yields in the search regions for the
  inclusive and $\tau$~\subref{figure4a} and high-\pt dilepton~\subref{figure4b} selections. For the
  inclusive selections, the results of method (B) are compared with those from method (A1) in the
  left and right bar for each channel, respectively. For the high-\pt selections, the results of
  method (A2) are compared with those from method (A1) in the left and right bar for each channel,
  respectively. Predictions for events with one and two fakes (prompt-fake and fake-fake),
  contributions from simulated backgrounds (SS prompt-prompt), and those from events with a lepton
  charge misreconstruction (OS prompt-prompt) are reported separately.}
\label{fig:figure4}
\end{figure*}



\section{Interpretation of Results for New Physics Models}

It is necessary to convey the information shown in Table~\ref{table1} in a form that can be used to
test a variety of specific physics models. This is done by using generator-level simulation studies
as approximations for the models. This was shown to be sufficiently precise to reproduce constraints
on new physics models that would otherwise require the full CMS detector simulation.

\begin{figure*}[hbtp]
\centering
\subfloat[][]{
  \includegraphics[width = 80mm]{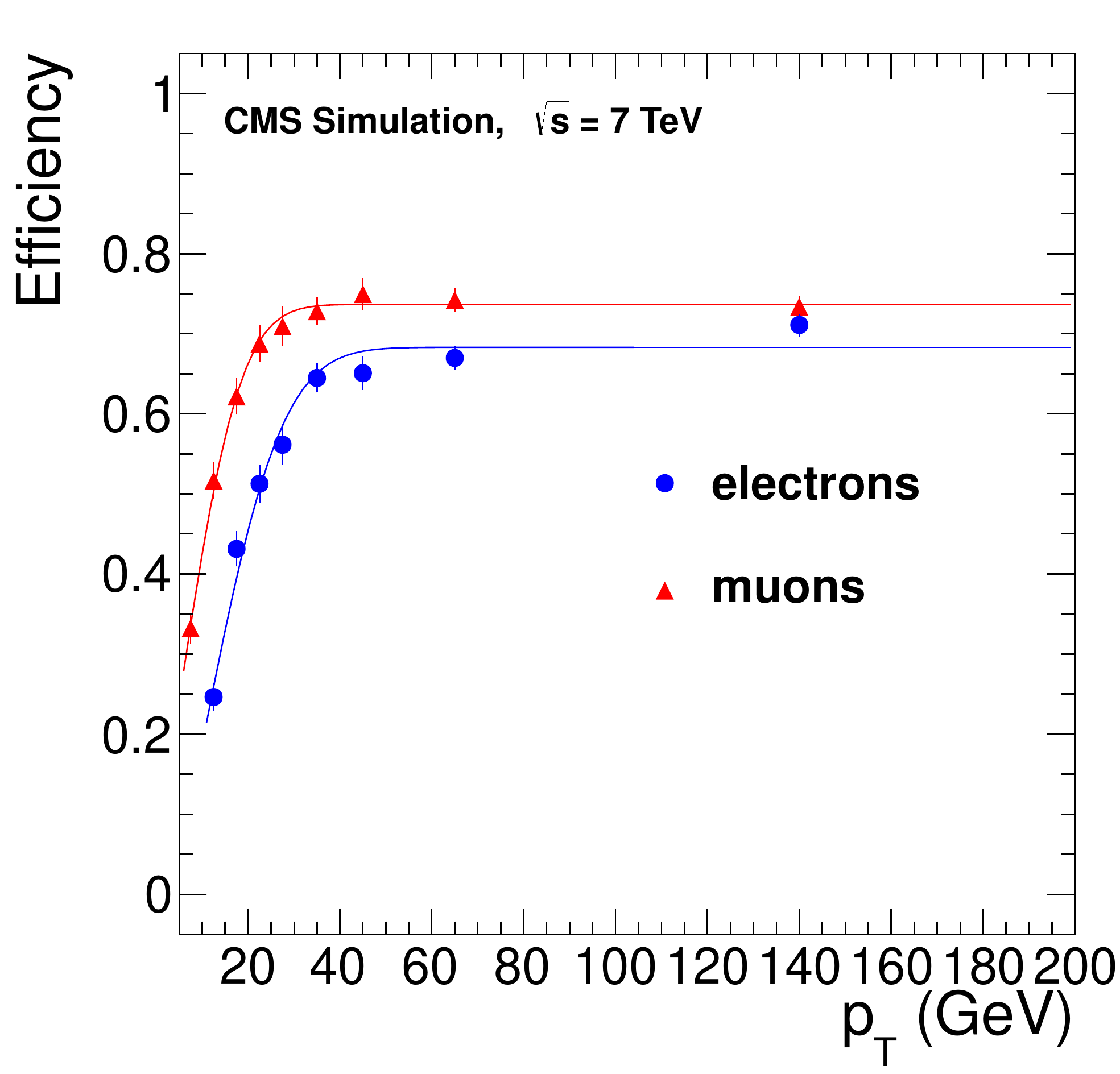}
  \label{figure16a}
}
\subfloat[][]{
  \includegraphics[width = 80mm]{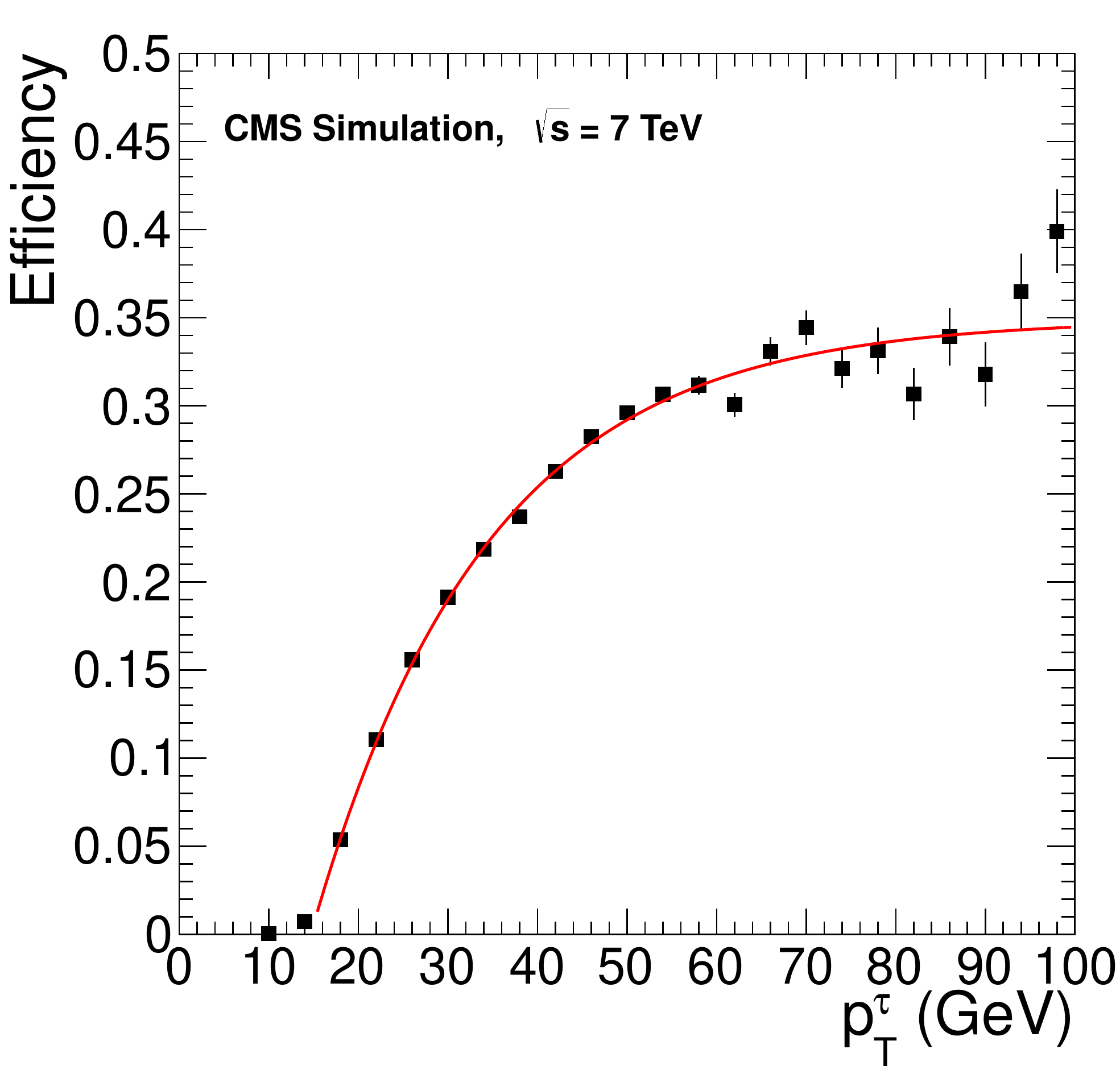}
  \label{figure16b}
}
\caption{Electron~\subref{figure16a} and muon~\subref{figure16b} selection efficiency as a function
  of~\pt, estimated in simulation LM6 benchmark point and corrected for simulation-to-data scale
  factors.}
\label{fig:figure16}
\end{figure*}

\begin{figure*}[hbtp]
\centering
\subfloat[][]{
  \includegraphics[width = 80mm]{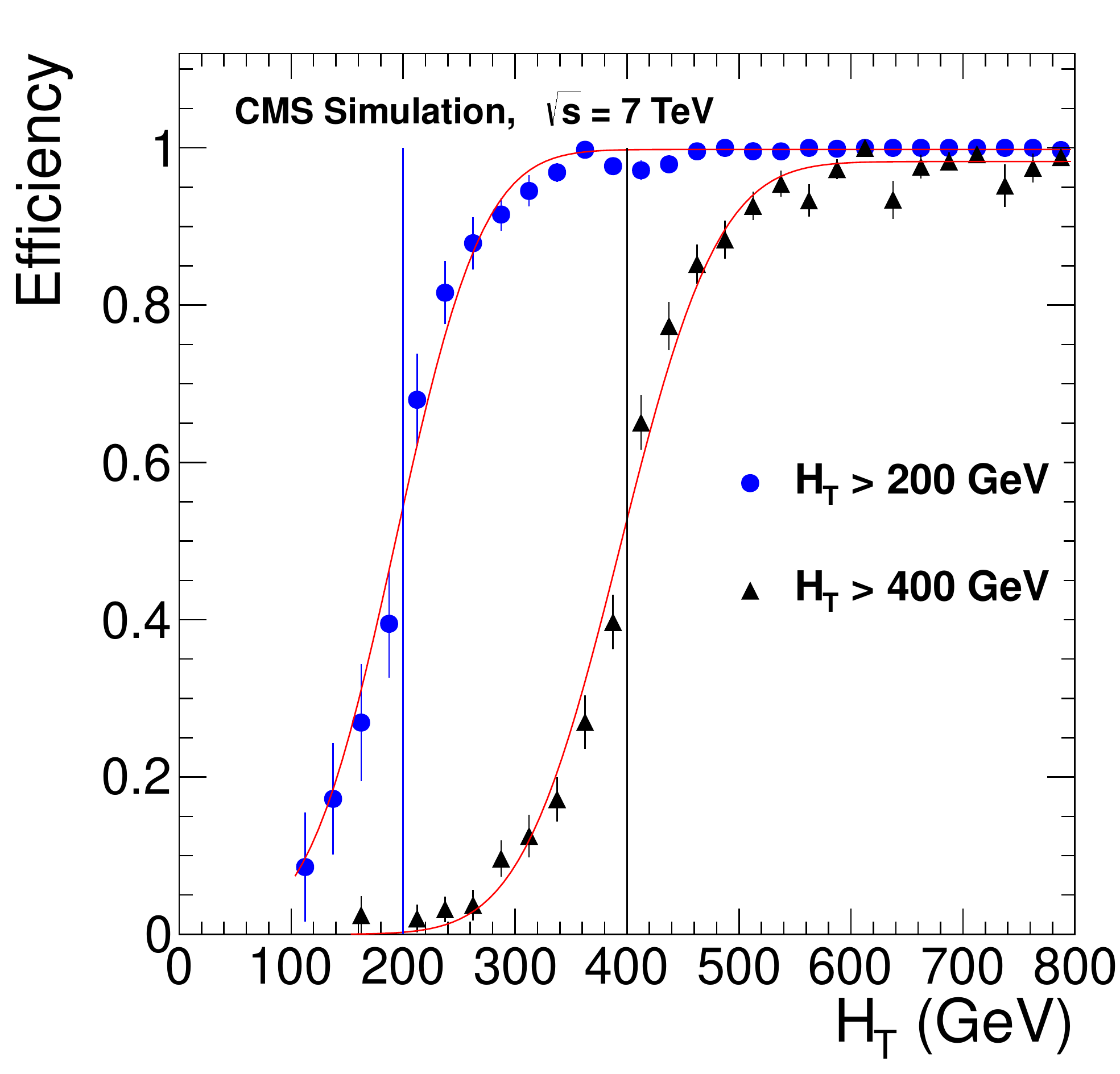}
  \label{figure17a}
}
\subfloat[][]{
  \includegraphics[width = 80mm]{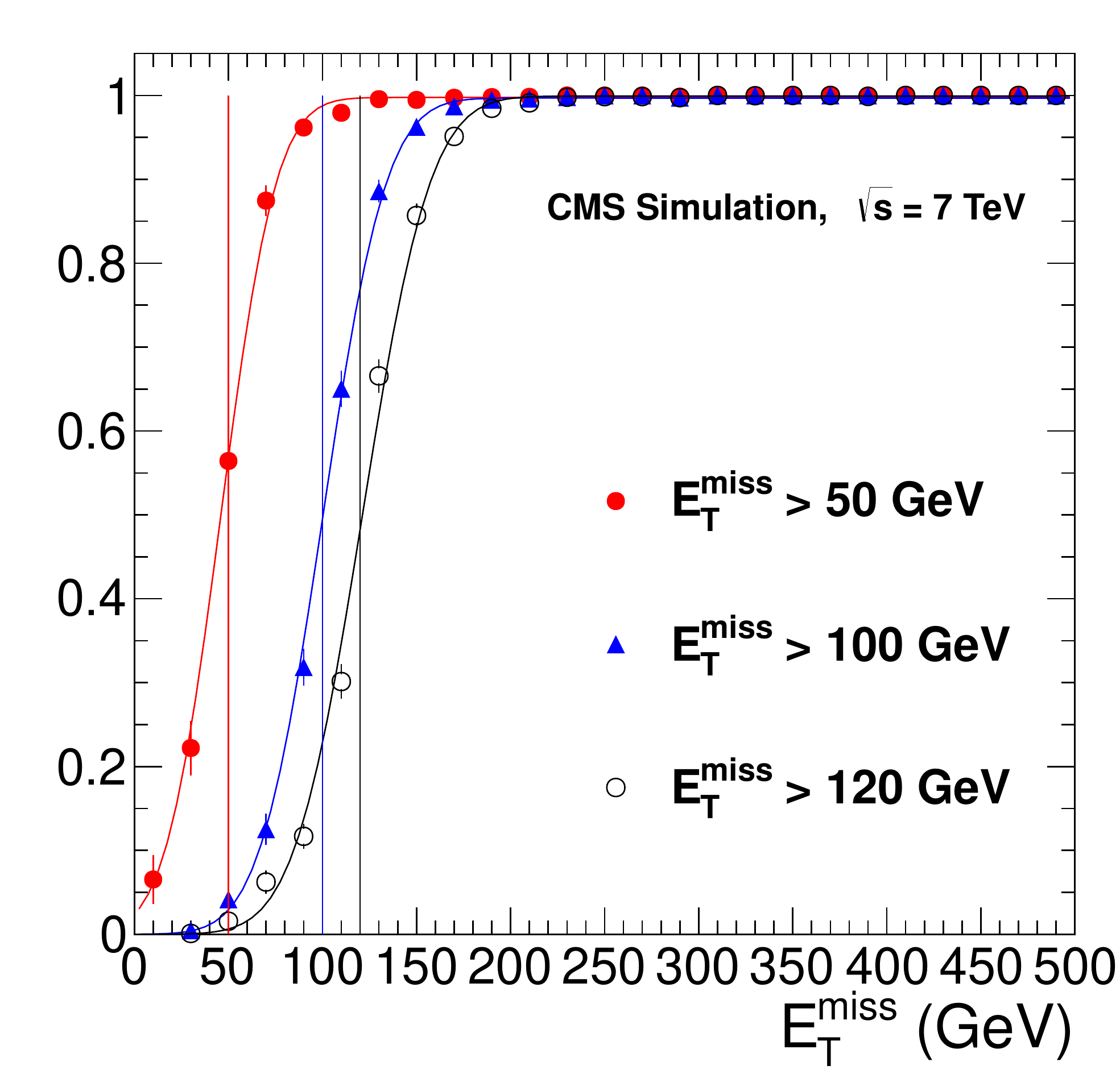}
  \label{figure17b}
}
\caption{Efficiency for an event to pass a given
  reconstructed~\hT~\subref{figure17a} and~\met~\subref{figure17b} threshold as a function of
  generator level~\hT and~\met. The curves are shown for~\met thresholds of 50, 100, and 120~\gev;
  the thresholds for~\hT are 200 and 400~\gev.}
\label{fig:figure17}
\end{figure*}

We take as a benchmark point the low-mass CMSSM parameter point LM6. The efficiency of the lepton
selection is shown in Figure~\ref{fig:figure16} as a function of~\pt, and the efficiency of the~\hT
and~\met selection is shown in Figure~\ref{fig:figure17} as a function of the reconstructed~\hT
and~\met, respectively. This efficiency dependence can be parametrized by error functions to
determine the overall selection efficiency for new physics signals.

\begin{figure}[!ht]
\centering
\includegraphics[width = 80mm]{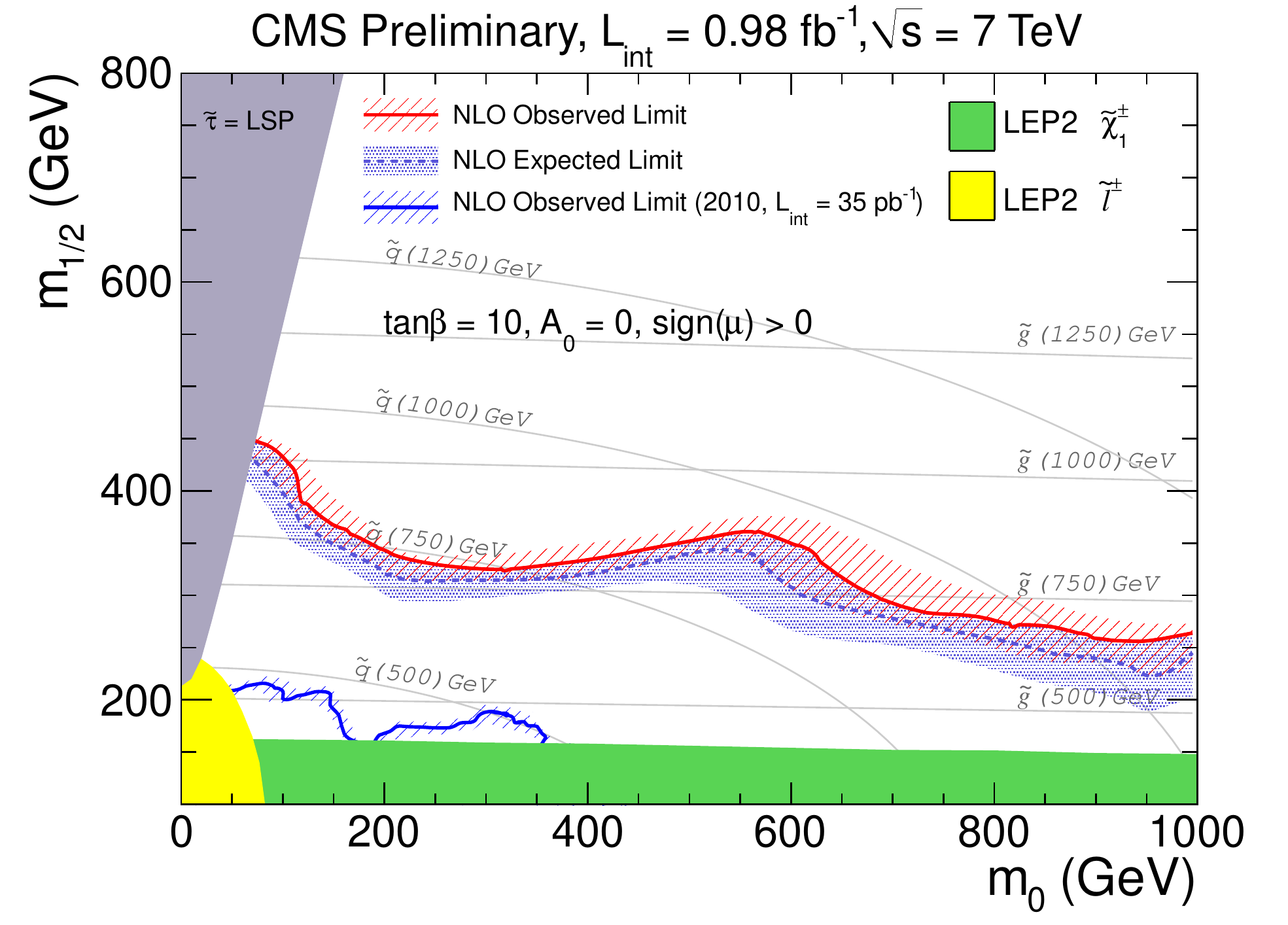}
\caption{Exclusion region in the CMSSM corresponding to the observed upper limit of 3.0 events in
  the search region 1 of the high-\pt dilepton selections. The result of the previous analysis is
  shown to illustrate the improvement since 2010.} \label{figure5}
\end{figure}

In order to provide a reference for other SUSY searches, the results are interpreted in the context
of the CMSSM model. The observed upper limits on the number of signal events shown in
Table~\ref{table1} for the high-\hT, high-\met search region of the high-\pt dilepton baseline
selection are compared to the expected number of events in the CMSSM model in a plane of ($m_{0},
m_{1/2}$) for $\tan\beta = 10$, $A_{0} = 0$, and $\mu > 0$. All points on this plane with mean
expected values in excess of this limit are interpreted as excluded at the 95\% CL. This excluded
region is shown in Figure~\ref{figure5}. The shaded region represents the uncertainty on the
position of the limit due to uncertainty in the production cross section of CMSSM.

\section{Summary and Conclusions}

A search for new physics was performed in the same-sign dilepton channel, including final states
with electrons, muons, and taus. All major background sources were estimated directly from data.
Events with a single fake lepton were found to be the dominant background in all channels of the
search with the exception of the $\tau\tau$ channel, where two fake $\tau$ leptons were the primary
background contribution.

No evidence was seen for an excess over the background prediction. We therefore set 95\% CL upper
limits on the number of signal events within $|\eta| < 2.4$ with 0.98~\fbinv of data. These limits
are reported as an exclusion curve in CMSSM parameter space.


\pagebreak[4]


\begin{thebibliography}{9}   

\bibitem{susyLikeSign}
  R. Barnett, J. Gunion, and H. Haber, ``Discovering supersymmetry with like sign dileptons'',
  \emph{Phys. Lett.} \textbf{B 315} (1993) 349.

\bibitem{bosonicSusy}
  H. Cheng, K. Matchev, and M. Schmaltz, ``Bosonic supersymmetry? Getting fooled at the CERN LHC'',
  \emph{Phys. Rev.} \textbf{D 66} (2002) 056006.

\bibitem{particleDarkMatter}
  G. Bertone, D. Hooper, and J. Silk, ``Particle dark matter: Evidence, candidates and
  constraints'', \emph{Phys. Rept.} \textbf{405} (2005) 279.

\bibitem{elecCand}
  CMS Collaboration, ``Electron reconstruction and identification at $\sqrt{s} = 7~\tev$'',
  \emph{CMS Physics Analysis Summary} \textbf{CMS-PAS-EGM-10-004} (2010).

\bibitem{muonCand}
  CMS Collaboration, ``Performance of muon identification in $pp$ collisions at $\sqrt{s} =
  7~\tev$'', \emph{CMS Physics Analysis Summary} \textbf{CMS-PAS-MUO-10-002} (2010).

\bibitem{tauCand}
  CMS Collaboration, ``Performance of tau reconstruction algorithms in 2010 data collected with
  CMS'', \emph{CMS Physics Analysis Summary} \textbf{CMS-PAS-TAU-11-001} (2011).

\bibitem{particleFlow}
  CMS Collaboration, ``Particle-Flow Event Reconstruction in CMS and Performance for Jets, Taus,
  and~\met'', \emph{CMS Physics Analysis Summary} \textbf{CMS-PAS-PFT-09-001} (2009).

\bibitem{constrainedMinSusy}
  G. Kane et al., ``Study of constrained minimal supersymmetry'', \emph{Phys. Rev.} \textbf{D 49}
  (1994), no. 11, 6173.

\bibitem{confidenceLimit}
  Particle Data Group Collaboration, ``Review of particle physics'', \emph{J. Phys.} \textbf{G 37}
  (2010), 075021.




\end{thebibliography}

\end{document}